%% file: paper.tex
\documentclass[10pt,sigconf]{acmart}
\usepackage{graphicx} 

\usepackage{subcaption}
\usepackage{filecontents}
\usepackage{nameref}
\usepackage{tikz}
\usepackage{amsmath}

\usepackage{titlesec}
\makeatletter 
\newcommand\semilarge{\@setfontsize\semilarge{11}{12}}
\makeatother
\titleformat*{\subsection}{\semilarge\bfseries}

\usepackage{enumitem}
\usepackage{pbox}
\usepackage{multirow}
\usepackage{tabularx}
\usepackage{makecell}
\usepackage{mathtools}
\usepackage{caption}
\usepackage{float}
\usepackage{booktabs}
\usepackage[linesnumbered,ruled,vlined]{algorithm2e}
\usepackage{algorithm2e}
\usepackage{xcolor}
\usepackage{nameref}
\usepackage{wrapfig}
\usepackage{xspace}
\usepackage{url}
\usepackage{pifont}

\usepackage{breakurl}

\newcommand{\parab}[1]{\vspace{0.05in}\noindent\textbf{#1}}

\renewcommand\footnotetextcopyrightpermission[1]{}
\setcopyright{none}

\begin{document}
\title[GeoTrace]{Leveraging Traceroute Inconsistencies to Improve IP Geolocation}
\author{Alagappan Ramanathan, Sangeetha Abdu Jyothi}
\affiliation{%
\institution{University of California, Irvine}
\country{USA}
}

\input{sections/abstract}

\maketitle

\input{sections/intro}
\vspace{-2mm}
\input{sections/related}
\vspace{-3mm}
\input{sections/design}
\vspace{-2mm}
\input{sections/results}
\vspace{-3mm}
\input{sections/conclusion}

\bibliographystyle{ACM-Reference-Format}

\bibliography{paper}

\end{document}

%% file: sections/abstract.tex
\begin{abstract}
    Traceroutes and geolocation are two essential network measurement tools that aid applications such as network mapping, topology generation, censorship, and Internet path analysis. However, these tools, individually and when combined, have significant limitations that can lead to inaccurate results. Prior research addressed specific issues with traceroutes and geolocation individually, often requiring additional measurements. In this paper, we introduce GeoTrace, a lightweight tool designed to identify, classify, and resolve geolocation anomalies in traceroutes using existing data. GeoTrace leverages the abundant information in traceroutes and geolocation databases to identify anomalous IP addresses with incorrect geolocation. It systematically classifies these anomalies based on underlying causes—such as MPLS effects or interface discrepancies—and refines their geolocation estimates where possible. By correcting these inaccuracies, GeoTrace enhances the reliability of traceroute-based analyses without the need for additional probing. Our work offers a streamlined solution that enhances the accuracy of geolocation in traceroute analysis, paving the way for more reliable measurement studies.
\end{abstract}

%% file: sections/intro.tex
\section{Introduction}

Accurately mapping the structure and behavior of the Internet is crucial for numerous applications, from optimizing network performance and enhancing user experience to enforcing regulatory policies and combating security threats. Traceroute and IP geolocation are fundamental tools that researchers and network operators employ to map network paths and associate IP addresses with physical locations. These tools facilitate tasks such as topology generation~\cite{mtgnautilus, mtg2, mtg3}, censorship analysis~\cite{censor1, censor2}, routing path assessment~\cite{rpa1, rpa2, rpa3}, anomaly detection, and more.  

Despite their importance, traceroute and IP geolocation face inherent challenges that can lead to significant inaccuracies. Traceroute measurements are often obscured by factors such as Multiprotocol Label Switching (MPLS) tunnels and interface address variability. These issues can distort the measurements for reported paths, making it difficult to accurately interpret network routes. On the other hand, IP geolocation techniques often suffer from substantial inaccuracies, especially at finer granularities such as city or regional levels. Limitations of current geolocation techniques, along with inconsistencies in commercial geolocation databases, contribute to these errors.

When these challenges intersect, they compound inaccuracies in network analyses. Misinterpretations arising from faulty traceroute data combined with incorrect geolocations can lead to flawed network maps, misjudged performance metrics, and incorrect assumptions about data flow and jurisdiction. For example, in censorship analysis, such inaccuracies might obscure the true path of data through restrictive regions, leading to misunderstandings of censorship mechanisms and potentially ineffective countermeasures.

Existing solutions~\cite{mplsdonnet, iavhyun} typically address specific issues in isolation and often require additional active measurements or complex inference models. These methods can be resource-intensive, impose significant overhead, and do not scale well for large datasets. This gap highlights the need for a lightweight, scalable solution that systematically identifies and corrects anomalies in IP geolocation using only existing data.

In this paper, we introduce GeoTrace, a novel tool designed to identify, classify, and rectify anomalous geolocations of IP addresses based solely on data collected from traceroutes. GeoTrace leverages patterns and inconsistencies inherent in traceroute outputs to detect anomalies without the need for additional measurements or complex inference techniques. By focusing on the relationships between IP addresses and their immediate neighbors in a large corpus of traceroutes and employing an iterative refinement process, our approach effectively identifies anomalous IPs while simultaneously refining geolocation estimates for non-anomalous IPs.

GeoTrace addresses these inaccuracies by classifying anomalous IPs into two categories: MPLS-Affected IPs and Interface-Affected IPs. MPLS-Affected IPs are challenging to geolocate accurately due to uniform RTTs within MPLS tunnels, so GeoTrace identifies and flags them accordingly. For Interface-Affected IPs, influenced by interface address variability or database inaccuracies, GeoTrace refines their geolocations using techniques inspired by constraint-based geolocation methods. By leveraging accurately geolocated non-anomalous IPs as virtual vantage points — termed anchor IPs — GeoTrace estimates the locations of these anomalous IPs without additional measurements.

After identifying and classifying anomalous IP addresses, we analyze patterns and underlying trends associated with these anomalies. Towards that, we evaluated GeoTrace using real-world traceroute data comprising approximately 234,000 unique IPv4 addresses from seven million traceroutes. About 5.4\% of IPs were tagged as anomalous. GeoTrace effectively corrected the geolocations of all Interface-Affected IPs, enhancing data reliability. Compared to traditional speed-of-light validation methods—where only 30\% of IPs had a single geolocation cluster—GeoTrace achieved this for nearly 60\% of IPs, significantly reducing ambiguity without additional measurements. Our analysis also revealed systemic patterns, with geolocation databases often misassigning IPs to certain regions and major ASes exhibiting higher counts of anomalous IPs. Notably, around 30\% of the corrected IPs had country-level discrepancies with geolocation databases, indicating significant inaccuracies at even coarse granularities. These findings highlight GeoTrace's capability to enhance geolocation accuracy and uncover underlying issues in network measurements.

In summary, we make the following contributions
\begin{itemize}[leftmargin=*,nolistsep]
    \item We develop GeoTrace, a tool to systematically identify anomalous geolocations in traceroute data
    \item We propose methods to classify detected anomalies and apply corrections to improve geolocation accuracy, without additional active measurements, making it resource-efficient and scalable for large traceroute datasets.
    \item By analyzing the corrected anomalies, we observe patterns and trends that provide deeper insights into the prevalence and causes of inaccuracies.
\end{itemize}

%% file: sections/related.tex
\section{Related Work}

Traceroute and IP geolocation are fundamental tools in network measurement studies, employed for mapping and topology generation~\cite{mtgnautilus, mtg2, mtg3}, censorship analysis~\cite{censor1, censor2}, and routing path assessment and anomaly detection~\cite{rpa1, rpa2, rpa3}, amongst others. However, each tool presents inherent challenges and limitations. When combined, these challenges compound, leading to inaccuracies and misinterpretations. Previous studies often overlook these issues, underestimate their impact, or rely on expensive active measurements to generate or validate findings. This section reviews the challenges identified in prior research, focusing on geolocation inaccuracies, the effects of MPLS tunneling on traceroute measurements, interface address variability, and issues introduced by /31 subnets on point-to-point links. 

\parab{(i) Geolocation inaccuracies at broader granularities:} IP geolocation maps IP addresses to physical locations, with research divided into latency-based techniques~\cite{latency1, latency2, latency3, latency4, latency5, latency6}, DNS-based approaches~\cite{dns1, dns2, dns3, dns4}, and statistical methods~\cite{stats1, stats2, stats3, stats4}. Several commercial geolocation databases also exist, such as MaxMind~\cite{maxmind}, IPinfo~\cite{ipinfo}, and NetAcuity~\cite{netacuity}. Despite extensive research and available databases, geolocation remains an active field due to persistent inaccuracies. Studies evaluating geolocation methods highlight substantial inaccuracies~\cite{geostudy1, geostudygharaibeh, geostudy3, geostudydarwich}. Gharaibeh et al.~\cite{geostudygharaibeh} evaluated public and commercial router geolocation databases, revealing substantial inaccuracies at both city and country levels. Some databases exhibited accuracy as low as 33\% at the country level, with the best performing around 75\% for certain countries like France and Singapore. Inaccuracies are especially pronounced near regional borders and AS boundaries. Similarly, Darwich et al.~\cite{geostudydarwich} assessed high-performing active measurement methods and found that none achieved satisfactory accuracy and coverage. These findings highlight challenges in geolocation accuracy, affecting applications requiring precise location data. 

\parab{(ii) Traceroute inaccuracies and complexities:} Traceroute records the route packets take to reach a destination, aiding in network mapping. However, traceroute measurements can be compromised due to various factors, leading to incorrect inferences about network topology and performance. Two significant challenges affecting traceroute accuracy are the effects of MPLS tunneling and interface address variability, which we expand on below.

\parab{(ii) (a) Effects of MPLS Tunneling on Traceroutes:} MPLS enhances network traffic flow by establishing label-switched paths for packets. While beneficial for performance, MPLS introduces complexities in interpreting traceroute data, as it can obscure the true packet path, affecting hop counts and RTT measurements. Though the Time-to-Live (TTL) propagation feature and RFC 4950~\cite{rfc4950} introduced ICMP extensions to include MPLS labels to aid identification, adoption is limited, reducing utility in network analysis. Donnet et al.~\cite{mplsdonnet} classified MPLS tunnels into explicit, implicit, opaque, and invisible types. Explicit and implicit tunnels cause nodes within the tunnel to have similar RTTs (RTT of the tunnel exit point), while opaque and invisible tunnels result in missing IPs within the tunnel, complicating path reconstruction and leading to incorrect inference.

Sommers et al.~\cite{mplssommers} examined MPLS deployments, identifying a significant presence of MPLS tunnels in traceroute paths using MPLS labels in traceroutes and Bayesian inference to detect explicit and implicit tunnels, respectively. Donnet et al.~\cite{mplsdonnet} proposed methods to identify implicit tunnels via targeted measurements. Vanaubel et al.~\cite{mplsvanaubel} studied invisible MPLS tunnels, presenting techniques to identify them. These studies demonstrate that MPLS tunneling is prevalent and significantly impacts traceroute interpretations. 

\parab{(ii) (b) Interface Address Variability in Traceroutes:} Traceroute relies on ICMP Time Exceeded messages from routers, ideally containing the ingress interface IP where the packet arrives. However, per RFC 1812~\cite{rfc1812}, routers respond with the interface over which the ICMP message is transmitted. Additionally, routers may be configured to use different IPs, leading to variability such as egress interfaces, loopback addresses, or off-path interfaces in traceroutes. This variability can lead to incorrect topology inference and complicates geolocation, especially near AS boundaries or country borders where accurate location is crucial. Extended ICMP messages specified in RFC 4884~\cite{rfc4884} and RFC 5837~\cite{rfc5837} allow routers to include ingress interface information, but limited adoption reduces effectiveness. Several studies address this challenge. Hyun et al.~\cite{iavhyun} examined third-party addresses in traceroute paths. Luckie and Claffy~\cite{iavluckie} proposed using the IP Timestamp option to detect such addresses, while Marchetta et al.~\cite{iavmarchetta} and Marder et al.~\cite{iavmarder} advanced techniques to identify interface variability. These methods often require additional probing or complex inference models, increasing measurement overhead. 

\parab{(iii) Challenges with /31 Subnets on P2P Links:} RFC 3021~\cite{rfc3021} permits /31 prefixes on point-to-point (P2P) links to conserve IPv4 addresses. While efficient, it introduces challenges in traceroute outputs. Hu et al.~\cite{31he} measured route asymmetry considering /30 and /31 subnets, finding about 10\% asymmetry in commercial Internet links.

\parab{Impact on Network Analysis:} The intersection of these challenges amplifies potential errors in inference based on traceroutes and geolocations, hindering reliability. For instance, MPLS tunnels might result in hops with similar RTTs, obscuring the true path. Combined with interface variability reporting off-path interfaces and inconsistent geolocation for these interfaces near AS or country boundaries, ambiguity compounds. Researchers have proposed methods addressing specific traceroute and geolocation challenges, often focusing on individual issues and requiring additional measurements or complex techniques. Studies on MPLS tunnels~\cite{mplsdonnet, mplssommers, mplsvanaubel, mpls4, mpls5, mpls6} and interface variability~\cite{iavhyun, iavluckie, iavmarchetta, iavmarder} provide insights but can be resource-intensive and may not scale to large datasets. They may also lack the ability to address overlapping challenges simultaneously. Thus, there is a need for lightweight tools that can systematically identify and rectify anomalous geolocations without additional measurements.

\vspace{-2mm}

%% file: sections/design.tex
\section{Design}

\subsection{Identifying Anomalous IP Addresses}

In this section, we present GeoTrace's methodology for identifying anomalous IP addresses within traceroutes while simultaneously refining their geolocation estimates. GeoTrace uses an iterative mechanism that relies solely on existing traceroute data from measurement platforms to systematically reduce ambiguity in IP geolocations and detect anomalies without the need for additional measurements. 

\parab{Geolocation Aggregation from Multiple Databases:} GeoTrace begins by extracting unique IP addresses from the collected traceroutes and obtaining their geolocations from multiple IP geolocation databases. Prior research~\cite{mtgnautilus} has demonstrated that employing multiple databases enhances coverage and accuracy due to the varying data sources and methodologies each database uses. Hence GeoTrace collects geolocation data from eight different databases~\cite{ip2location, ipinfo, db-ip, ipregistry, ipgeolocation, ipapi, ipapico, ipdata}, commonly used in previous studies. To manage discrepancies among databases, GeoTrace clusters the geolocations that map to the same city, which reduces the number of location candidates for each IP address and simplifies subsequent analysis. 

\parab{Ideal Approach and Its Computational Challenges:} Ideally, to determine the correct geolocation or identify anomalies for a specific IP address involved in multiple traceroutes, one might exhaustively evaluate all possible combinations of its geolocation candidates. For each traceroute, assuming accurate geolocations for the other IPs, one would assess how well each candidate location for the target IP aligns with the observed round-trip times (RTTs). The candidate whose geographical distances consistently correlate with RTTs across the majority of traceroutes would be considered the most plausible. If none of the candidates align adequately with the RTTs, it suggests that the IP's geolocation is anomalous.

However, this exhaustive approach is computationally infeasible due to the exponential growth in the number of path combinations. Despite clustering geolocations, IP addresses often have multiple potential locations, especially near autonomous system (AS) boundaries or country borders. For example, a traceroute with several IPs having multiple location candidates can result in evaluating thousands or even hundreds of thousands of possible paths. This combinatorial explosion renders exhaustive analysis impractical.

\subsubsection{Iterative Neighbor-Based Evaluation}
To overcome the computational challenge, GeoTrace employs an iterative approach that focuses on an IP address's local neighborhood, i.e., its immediate neighbors comprising the preceding and following hop in a traceroute. By limiting the evaluation to neighboring IP pairs rather than entire paths, we significantly reduce computational complexity. 

We use a scoring mechanism to assess the suitability of each geolocation candidate for an IP address. For each IP and its neighbor, GeoTrace evaluates the feasibility of their geolocation combinations by comparing the difference in their RTTs to the geographical distance between their candidate locations, calculated using the Haversine distance~\cite{haversine_distance}. To accommodate variability in network conditions and latency discrepancies, we introduce a dynamic deviation allowance. This allowance is calculated as a percentage (10\%) of the sum of the RTTs of the two hops, ensuring our model adapts to inherent variations in traceroute measurements. If the RTT difference and the geographical distance align within the deviation allowance, the combination is considered feasible.

After evaluating all candidate combinations for an IP and its neighbors, GeoTrace computes a performance ratio for each geolocation candidate, defined as the number of successful (feasible) evaluations divided by the total number of evaluations for that candidate. GeoTrace then retains only those geolocation candidates whose performance ratios are within a specific threshold (90\%) of the best-performing candidate, effectively pruning unlikely locations and focusing on the most promising options.

The iterative process continues, with each iteration potentially refining the geolocation options for IPs based on updated information from their neighbors. The process repeats until changes in geolocation candidates become negligible between iterations, indicating that stable and accurate location estimates have been reached. This convergence ensures that the solution becomes progressively more refined and reliable with each iteration.

GeoTrace's approach effectively mitigates the impact of network anomalies and error propagation. By leveraging data from multiple traceroutes, GeoTrace averages out transient network conditions affecting measurements, reducing the influence of anomalies on the final geolocation estimates. Redundant paths serve as cross-validation, enabling the identification and correction of outliers. The iterative nature of the process allows stable geolocation estimates for some IPs to help refine those for neighboring IPs over successive iterations. A key outcome of GeoTrace's methodology is the identification of anomalous IP addresses. After convergence, GeoTrace evaluates the performance ratios of the remaining geolocation candidates for each IP. An IP is tagged as anomalous if its performance ratios are consistently low across all geolocation options, or if it shows a significant discrepancy in alignment with either its previous or next hop, but not both. This tagging highlights potential irregularities and facilitates further study of these IPs, which is crucial for network diagnostics, security analyses, and improving the reliability of geolocation data. Moreover, GeoTrace offers dual benefits: it effectively detects anomalous IP addresses and refines the geolocation estimates for non-anomalous IPs. By filtering and confirming the most plausible locations, GeoTrace enhances the overall fidelity of IP geolocation.

\vspace{-2mm}

\subsection{Resolving and Classifying Anomalous IP Addresses}

Building upon the identification of anomalous IP addresses in the previous stage, in this section, we introduce GeoTrace's methodology for resolving their geolocations where possible and classifying them based on the underlying causes of anomaly. This process not only refines geolocation estimates but also reduces false positives from the initial anomaly detection, enhancing the overall accuracy of network analyses.

GeoTrace categorizes anomalous IPs into two primary groups: \textit{MPLS-Affected IPs} and \textit{Interface-Affected IPs}. MPLS-Affected IPs are those impacted by MPLS tunnels, with similar RTTs for all hops within the tunnel, which is the RTT of the tunnel exit. In contrast, Interface-Affected IPs are influenced by factors such as interface address variability or inaccuracies in geolocation databases; however, their geolocations can be approximated more accurately.

GeoTrace leverages an approach inspired by active measurement geolocation techniques that have been adapted to operate without additional measurements to resolve the geolocations of anomalous IPs. Instead of relying on external vantage points, GeoTrace utilizes the accurately geolocated non-anomalous IPs identified earlier as virtual vantage points, referred to as \textit{anchor IPs}. For each anomalous IP address, GeoTrace examines all associated traceroutes and performs a bidirectional search along the network path to identify the closest anchor IPs on both sides. To determine the most suitable anchor IP for each anomalous IP in a given traceroute, GeoTrace selects the anchor IP with the smallest absolute RTT difference from the anomalous IP. These selected anchor IPs serve as reference points, analogous to vantage points in active measurement-based geolocation methods. The differences in RTTs between the anomalous IP and anchor IP are used as proxies for delay, providing insights into the geographical proximity between the IPs. 

However, since absolute RTT differences can be influenced by transient network conditions and may not always accurately reflect true proximity, GeoTrace employs two strategies to mitigate potential inaccuracies. First, by leveraging multiple traceroutes, GeoTrace reduces the impact of transient fluctuations by considering the median RTT difference for each anchor-anomalous IP pair. This statistical aggregation smooths out anomalies and provides a more stable estimate. Second, GeoTrace introduces a dynamic deviation allowance calculated as a percentage of the observed RTTs of the selected anchor IP. This allowance accounts for inherent variability in network measurements, ensuring that minor discrepancies do not lead to incorrect conclusions.

Before attempting to resolve the geolocations, GeoTrace filters out anomalous IPs likely impacted by MPLS tunnels. It analyzes the geographical distribution of the selected anchor IPs associated with each anomalous IP. If the anchor IPs are dispersed across multiple countries or continents—with no single country accounting for more than 95\% of them—the anomalous IP is classified as an MPLS-Affected IP. For the remaining anomalous IPs not classified as MPLS-affected, GeoTrace proceeds to refine their geolocations using a methodology inspired by constraint-based geolocation techniques. GeoTrace constructs buffer regions around the geolocations of the selected anchor IPs, utilizing the median RTT differences to define the sizes of these regions, where the anomalous IP could reside.

Recognizing that the exact overlap of all buffer regions is improbable due to measurement inaccuracies, GeoTrace aims to identify regions where multiple buffer zones converge, indicating a higher likelihood of the anomalous IP's true location. To facilitate efficient computation, GeoTrace employs geographic clusters known as 
\textit{city polygons}, representing major urban areas formed by clustering intersections of transportation infrastructures such as roads and railways.

Using spatial indexing techniques, GeoTrace quickly determines which city polygons have the highest count of overlapping buffer regions. The anomalous IP is then assigned the geolocation corresponding to the centroid of the city polygon with the maximal overlap. In cases where multiple city polygons exhibit equal maximum overlap, GeoTrace groups these regions based on proximity, merging those within a specified threshold distance (20 km) into clusters. If ambiguity persists even after increasing the threshold (up to 100 km), the anomalous IP is classified as an MPLS-Affected IP due to the inability to accurately pinpoint its location.

After resolving geolocations, GeoTrace compares the newly determined locations with the initial geolocations obtained from databases. If there is a significant discrepancy between the resolved location and the original database geolocation, the IP is classified as an Interface-Affected IP. Conversely, if the resolved location closely matches the original geolocation, GeoTrace considers the prior anomaly tagging as a false positive, often due to limited data, and updates its classification accordingly. 

By resolving geolocations and accurately classifying anomalous IPs, GeoTrace significantly improves the fidelity of network geolocation data. This methodological approach offers several advantages. It enhances geolocation accuracy by leveraging anchor IPs and employing statistical techniques, which refine the estimates for previously anomalous IPs. Furthermore, reducing false positives strengthens the reliability of network analyses, as only genuinely anomalous IPs are flagged. Through this seamless integration of anomaly identification, geolocation resolution, and classification, GeoTrace provides a robust framework for improving network mapping and analysis.

%% file: sections/results.tex
\section{Results and Analysis}

In this section, we present the results produced by GeoTrace and examine the patterns and trends demonstrated by the anomalous IP addresses identified and corrected by our methodology. Specifically, we address the effectiveness of GeoTrace in refining geolocation estimates and analyze the characteristics of the Interface-Affected IPs whose locations we corrected.

\parab{Experimental Setup:} To evaluate the performance of GeoTrace and identify patterns among anomalous IPs, we collected traceroute data from the RIPE Atlas measurements 5051 and 5151 for one day (March 5, 2024). This dataset comprised $\approx$ seven million traceroutes, involving 234,000 unique IPv4 IP addresses and 328,000 unique links. The substantial volume of data provides a robust foundation for assessing GeoTrace's capabilities in a real-world context.

\parab{GeoTrace Classification and Correction Analysis:} Applying GeoTrace's identification, classification, and correction processes to the traceroute data yielded significant insights. Of the 234,000 IP addresses analyzed, $\approx$ 5.4\% were tagged as anomalous, with an almost equal split between MPLS-Affected IPs and Interface-Affected IPs. The impact of anomalous IPs was more pronounced when considering links and traceroutes. $\approx$ 20\% of the links and 55\% of the traceroutes were affected by the presence of anomalous IPs. This substantial influence highlights how anomalies can propagate through network structures, potentially distorting analyses that rely on traceroute data. It is important to note that individual links or traceroutes could involve both MPLS-Affected and Interface-Affected IPs, leading to some overlap in the affected counts. 

Further analysis revealed that MPLS-Affected IPs had a greater impact on links and traceroutes compared to Interface-Affected IPs, as detailed in Table~\ref{tab:summary}. By employing GeoTrace's correction methodology, we successfully resolved the geolocations for all Interface-Affected IPs. Consequently, we corrected the links and traceroutes that were exclusively impacted by the presence of Interface-Affected IPs. Specifically, this correction accounted for $\approx$ 49\% of the impacted IPs, 40\% of the affected links, and 26\% of the impacted traceroutes. These results demonstrate GeoTrace's effectiveness in mitigating the influence of anomalies and enhancing the accuracy of network analyses.

\begin{table}[h]
    \centering
    \small
\begin{tabular}{lccc}
    \toprule
    \textbf{Category} & \textbf{IPs} & \textbf{Links} & \textbf{Traceroutes} \\
    \midrule
    \textbf{Total Elements} & 234K & 328K & 7.0M \\ \hline
    MPLS-Affected & 6.5K (2.8\%) & 41K (12.5\%) & 2.9M (41.7\%) \\ 
    Interface-Affected & 6.3K (2.6\%) & 29K (8.8\%) & 1.8M (25.7\%) \\ 
    \textbf{Total Affected} & 12.8K (5.4\%) & 68K (20.7\%) & 3.9M (55.7\%)  \\  \hline
    \textbf{Corrected} & 6.3K (49.2\%) & 27K (39.7\%) & 1M (25.6\%) \\ 
    \bottomrule
\end{tabular}
    \caption{\small IP, Link, and Traceroute Statistics. Corrected denotes the number of elements corrected by GeoTrace. Percentages for affected are with respect to total elements and those for corrected are with respect to the total affected.}
    \label{tab:summary}
\end{table}

\begin{figure}[h]
    \centering
    \begin{minipage}{0.45\columnwidth}
        \centering
        \includegraphics[width=\textwidth]{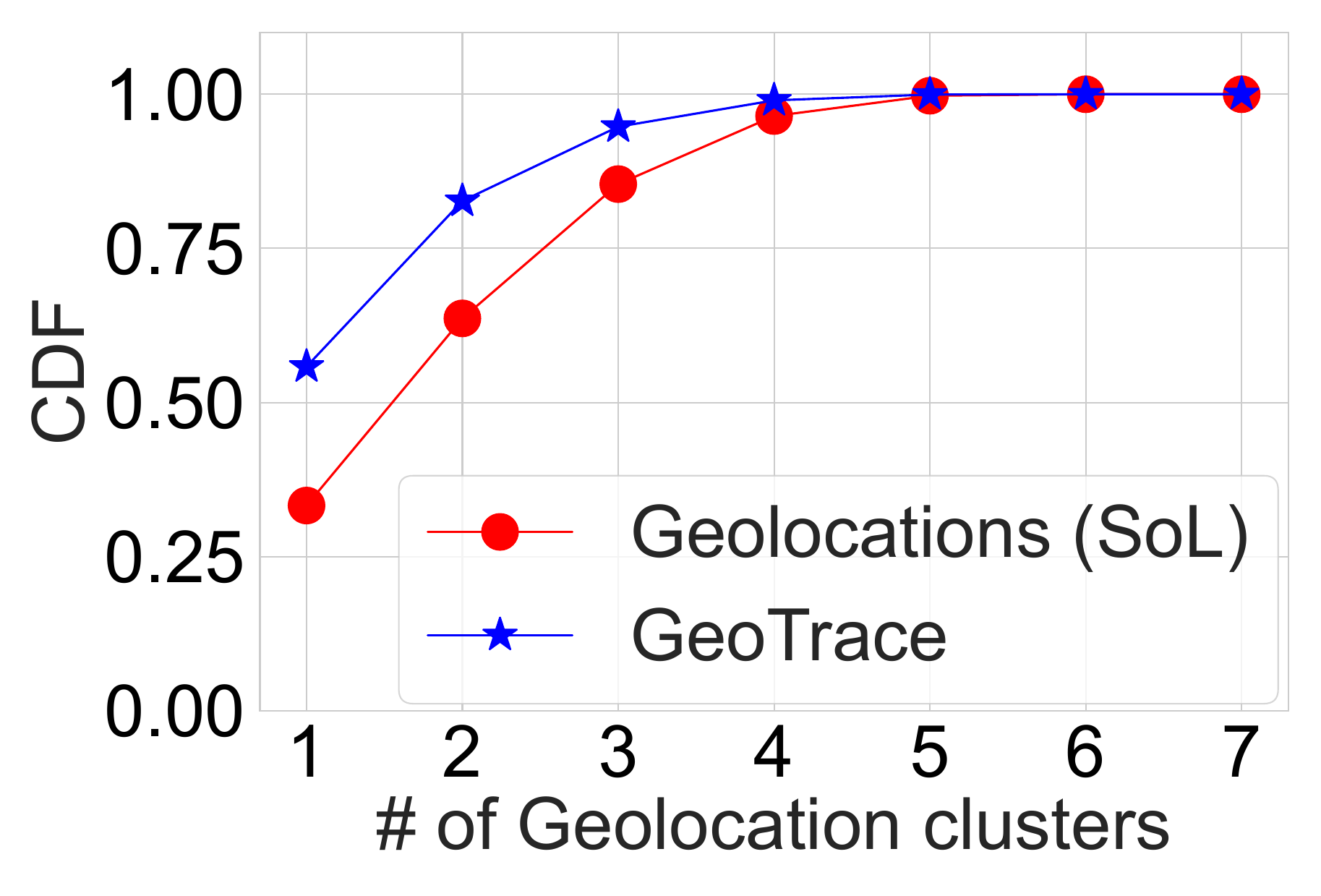}
        \caption{\small The distribution of unique geolocations per IP address. GeoTrace shifts geolocations closer to the ideal of 1 per IP address.}
        \label{fig:GeoTrace_improve}
    \end{minipage}\hfill
    \begin{minipage}{0.45\columnwidth}
        \centering
        \includegraphics[width=\textwidth]{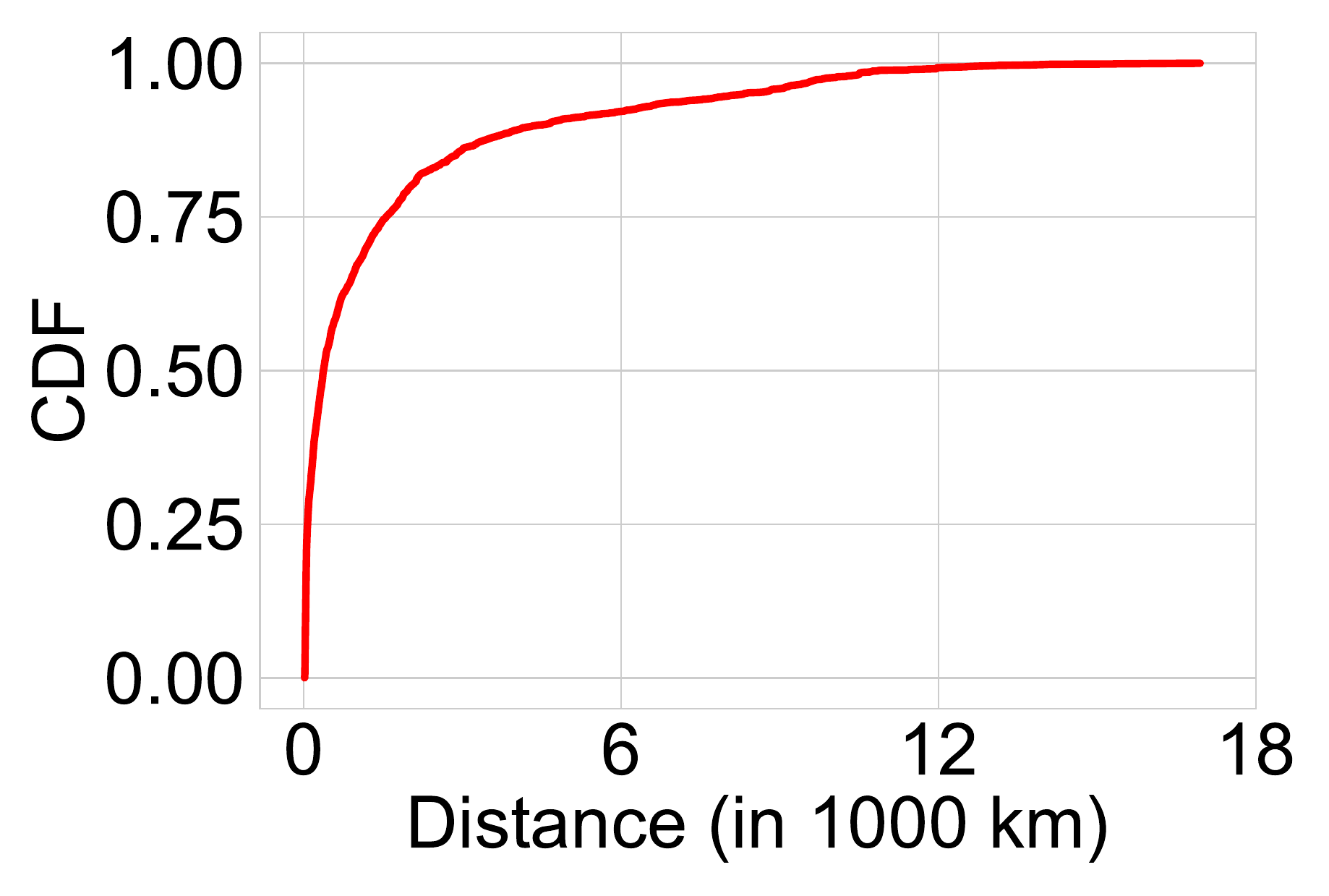}
        \caption{ \small  The distribution of distances from the geolocation (agreed by most sources) to resolved geolocations for Interface-Impacted IPs}
        \label{fig:distance}
    \end{minipage}
\end{figure}

\parab{Geolocation Refinement Performance:} Beyond identifying and correcting anomalous IPs, GeoTrace also refines the geolocation choices for non-anomalous IPs. To assess the improvement in geolocation refinement, we compared the geolocations derived by GeoTrace with those obtained using a common approach that clusters geolocations post Speed-of-Light (SoL) validation, as employed in several prior works. In an ideal scenario, each IP address should correspond to a single geolocation cluster, indicating consistent and accurate geolocation data. Using the SoL validation method, only about 30\% of the IP addresses had a single geolocation cluster. In contrast, GeoTrace achieved a significant improvement, with nearly 60\% of IP addresses having a single geolocation cluster. Moreover, $\approx$ 95\% of IP addresses had three or fewer geolocation clusters when processed with GeoTrace. These results, illustrated in Figure~\ref{fig:GeoTrace_improve}, highlight the substantial enhancement in geolocation accuracy achieved by GeoTrace without the need for additional measurements.

\parab{Analysis of Corrected IPs:} To delve deeper into the characteristics of the Interface-Affected IPs whose locations were corrected by GeoTrace, we examined the data along two key dimensions: distance distribution and country-level trends.

\parab{Distance Distribution between Resolved and Original Geolocations:} We first assessed the distance between the resolved locations provided by GeoTrace and the geolocations agreed upon by the majority of geolocation services for each IP address. Ideally, if the geolocation services were highly accurate, this distance would be negligible. However, as depicted in Figure~\ref{fig:distance}, only about 5\% of the corrected IPs had a distance of less than 20 km (identified as city radius in prior work) between the resolved location and the geolocation services' consensus. The average distance was $\approx$ 1,500 km, with the maximum discrepancy reaching up to 17,000 km. These findings indicate significant inaccuracies in the geolocation services' data and underscore the value of GeoTrace's corrections.

\begin{figure}[h]
    \centering
    \includegraphics[width=\columnwidth]{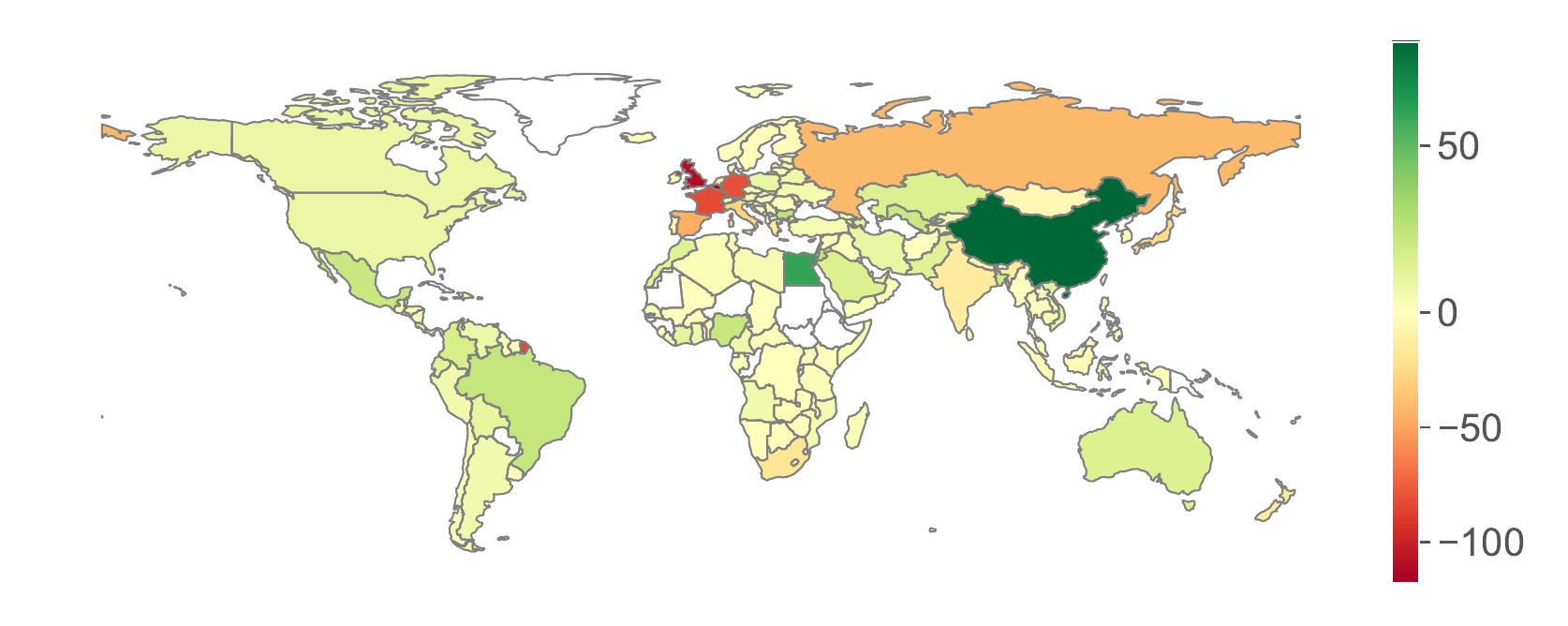}
    \caption{\small Heatmap of changes in the number of geolocations for Interface-Impacted IPs at the country level. }
    \label{fig:heatmap}
\end{figure}

\parab{Country-Level Trends:} Next, we examined the trends at the country level by comparing the counts of Interface-Affected IPs assigned to each country based on the geolocation services' consensus and the resolved locations computed by GeoTrace based on RTT measurements. We calculated the differences in these counts and visualized them in a heatmap, as shown in Figure~\ref{fig:heatmap}. A negative value indicates that more IP addresses were mapped to the country by geolocation databases than by GeoTrace, while a positive value suggests the opposite. Our analysis revealed notable discrepancies in certain regions. Countries in Western Europe, specifically the United Kingdom, France, and Germany, exhibited significant negative values. This suggests that geolocation databases inaccurately mapped several IP addresses to these European countries. Conversely, in China, geolocation services assigned fewer of these anomalous IP addresses compared to GeoTrace's RTT-based assessments, indicating underrepresentation. Further investigation showed that $\approx$ 1,900 IPs—representing about 30.36\% of the corrected IPs—had discrepancies at the country level between the geolocation databases and GeoTrace's resolved locations. These results highlight inaccuracies in geolocation services at the country level, consistent with observations from prior research.

%% file: sections/conclusion.tex
\section{Conclusion}

In this paper, we presented GeoTrace, a lightweight and scalable tool designed to enhance the accuracy of IP geolocation using only existing traceroute data. By systematically identifying, classifying, and correcting anomalous IP addresses, GeoTrace addresses the compounded inaccuracies that arise from factors such as MPLS tunnels and interface address variability. Our methodology leverages an iterative neighbor-based evaluation process and refines geolocation estimates without the need for additional active measurements. Through experiments using real-world traceroute datasets, GeoTrace demonstrated significant improvements in geolocation accuracy. We successfully corrected all Interface-Affected IPs, reducing ambiguity and enhancing data reliability. Our analysis uncovered systemic patterns of inaccuracies in geolocation databases, with approximately 30\% of corrected IPs exhibiting country-level discrepancies.